\documentclass[twocolumn,showpacs,nofootinbib,aps,superscriptaddress,prd]{revtex4}
\usepackage{graphicx,dcolumn,bm,amsfonts,amsmath,color,xcolor}
\usepackage[dvipdfm,colorlinks=true,citecolor=blue, linkcolor=blue,urlcolor=blue]{hyperref}
\usepackage{slashed}

\allowdisplaybreaks

\begin{document}

\title{Constraint of $\xi$-moments calculated with QCD sum rules on the pion distribution amplitude models}

\author{Tao Zhong\footnote{Corresponding author}}
\email{zhongtao1219@sina.com}
\address{School of physics, Henan Normal University, Xinxiang 453007, P.R. China}
\address{Department of Physics, Guizhou Minzu University, Guiyang 550025, P.R. China}
\author{Zhi-Hao Zhu}
\address{School of physics, Henan Normal University, Xinxiang 453007, P.R. China}
\author{Hai-Bing Fu}
\email{fuhb@cqu.edu.cn}
\address{Department of Physics, Guizhou Minzu University, Guiyang 550025, P.R. China}
\address{Department of Physics, Chongqing University, Chongqing 401331, P.R. China}

\date{\today}

\begin{abstract}
So far, the behavior of the pionic leading-twist distribution amplitude (DA) $\phi_{2;\pi}(x,\mu)$ $-$ which is universal physical quantity and enters the high-energy processes involving pion based on the factorization theorem $-$ has not been completely consistent. The form of $\phi_{2;\pi}(x,\mu)$ is usually described by phenomenological models and constrained by the experimental data of the exclusive processes containing pion or the moments calculated with the QCD sum rules and lattice QCD theory. Obviously, an appropriate model is very important for us to determine the exact behavior of $\phi_{2;\pi}(x,\mu)$. In this paper, by adopting the least squares method to fit the $\xi$-moments calculated with QCD sum rules based on the background field theory, we perform an analysis for several commonly used models of the pionic leading-twist DA in the literature, such as the truncation form of the Gegenbauer polynomial series, the light-cone harmonic oscillator model, the form from the Dyson-Schwinger equations, the model from the light-front holographic AdS/QCD and a simple power-law parametrization form.
\end{abstract}


\maketitle

\section{introduction}

Further testing and developing the standard model (SM) and exploring the evidence for the existence of new physics (NP) beyond SM are the most important research projects in the field of particle physics. The $B/D$ to light meson exclusive decay processes provide a good platform for these major research projects. The physical quantities of interest in these processes, such as decay amplitude, decay width, decay branching ratio, etc., can usually be expressed in the form of convolution of the distribution amplitudes (DAs) of the final state light mesons. Therefore, light meson DAs, especially the leading-twist DAs, are the key parameters for predicting these exclusive decay processes, and they are also the main error sources. Their accuracy is directly related to the theoretical prediction accuracy of these processes. The discussion on meson DAs began in 1980, when Lepage and Brodsky used collinear factorization to study the large momentum transition process~\cite{Lepage:1980fj}. In the following 40 years, the pionic leading-twist DA $\phi_{2;\pi}(x,\mu)$ attracted the most attention. However, people still do not have a completely consistent understanding of the pionic leading-twist DA's behavior.

The pionic leading-twist DA $\phi_{2;\pi}(x,\mu)$ describes the momentum fraction distributions of partons in pion in a particular Fock state, which is a universal nonperturbative object, and cannot be measured directly through experiments. In principle, the nonperturbative QCD methods should be used to study the pionic leading-twist DA. However, due to the difficulty of the nonperturbative QCD, $\phi_{2;\pi}(x,\mu)$ is usually studied by combining the nonperturbative QCD such as QCD sum rules and lattice QCD theory (LQCD) with the phenomenological model~\cite{Zhong:2014jla, Zhong:2021epq}, or by combining the phenomenological model with relevant experimental data~\cite{Wu:2012kw, Huang:2013gra, Huang:2013yya, Zhong:2015nxa}. Obviously, the information about the pionic leading-twist DA obtained from the nonperturbative methods and related experiments needs to be expressed by the phenomenological model. In literature, there are several phenomenological models to be commonly used to describe the pionic leading-twist DA behavior, such as the truncation form of the Gegenbauer polynomial series based on $\{C_n^{3/2}\}$-basis (TF model), the light-cone harmonic oscillator model (LCHO model)~\cite{Huang:1994dy, Huang:2004fn, Wu:2011gf, Wu:2012kw, Huang:2013gra, Huang:2013yya, Zhong:2015nxa, Zhong:2014jla, Zhong:2021epq} based on the Brodsky-Huang-Lepage (BHL) description\cite{BHL}, the form from Dyson-Schwinger equations (DSE model)~\cite{Chang:2013pq}, the model from light-front holographic AdS/QCD (AdS/QCD model)~\cite{Chang:2016ouf, Chang:2018aut} and the power-law parametrization form (PLP model)~\cite{RQCD:2019osh}. In addition, the pionic leading-twist DA has been studied with the light-front quark model~\cite{Choi:2007yu, Dhiman:2019ddr}, light-front constituent quark model~\cite{deMelo:2015yxk}, the nonlocal chiral-quark model (NLChQM) from the instanten vacuum~\cite{Nam:2006au}, the Nambu-Jona-Lasinio model~\cite{Praszalowicz:2001wy, Praszalowicz:2001pi, RuizArriola:2002bp}, or by taking the infinite-momentum limit for the quasidistribution amplitude within NLChQM and LQCD based on the large-momentum effective theory~\cite{Nam:2017gzm}.

As a first-principle method, however, the computation of LQCD is usually limited to the first few moments of $\phi_{2;\pi}(x,\mu)$~\cite{Martinelli:1987si, Daniel:1990ah, DelDebbio:2002mq, Gockeler:2005jz, Braun:2006dg, Donnellan:2007xr, Arthur:2010xf, Braun:2015axa, Bali:2017ude, Detmold:2021qln, RQCD:2019osh}. Due to the systematic error arising from the truncation of the high-dimensional condensation terms and the approximation of the spectral density based on the quark-hadron duality, the study on the pionic leading-twist DA with the QCD sum rules is also limited to the lowest moments~\cite{Chernyak:1981zz, Chernyak:1983ej, Huang:1984nc, Xiang:1984hw, Johnston:1986zi, Mikhailov:1988nz, Mikhailov:1991pt, Bakulev:2001pa, Khodjamirian:2004ga, Ball:2006wn, Bakulev:1998pf}. On the other hand, one can also abstract the information of pionic twist-2 DA via the exclusive processes involving pion, such as pion-photon transition form factor~\cite{Schmedding:1999ap, Bakulev:2002uc, Agaev:2005rc, Agaev:2010aq, Agaev:2012tm}, pion electromagnetic form factor~\cite{Braun:1999uj, Bijnens:2002mg, Agaev:2005gu}, $B/D\to \pi l\nu$ semileptonic decays~\cite{Ball:2005tb}, etc. Then, the information about pionic leading-twist DA obtained from the nonperturbative methods and relevant experimental data is usually incomplete or indirect. Therefore, the choice of phenomenological model is particularly important to accurately describe the behavior of pionic leading-twist DA.

Although the calculation of moments using nonperturbative methods is usually limited to the lowest orders, there are several works in the literature worth mentioning, which try to calculate enough moments to obtain more complete information of the pionic leading-twist DA. Such as the aforementioned DSE model, the first 50 $x$-moments are calculated by solving the corresponding gap and Bethe-Salpeter Equations~\cite{Chang:2013pq}. Very recently, to resolve the long-term limitation on applications of QCD sum rules, Li proposed a new dispersion relation for the moments of pionic leading-twist DA~\cite{Li:2022qul}, and obtained the first 18 Gegenbauer moments by solving the dispersion relation as an inverse problem in the inverse matrix method~\cite{Li:2021gsx}. In Ref.~\cite{Zhong:2021epq}, we studied pionic leading-twist DA with QCD sum rules in the framework of background field theory~\cite{Huang:1989gv}. We clarified the fact that the sum rule of the zeroth $\xi$-moment of pionic leading-twist DA cannot be normalized in the whole Borel parameter region, and proposed a new sum rule formula for the $n$th $\xi$-moments. This new formula can reduce the systematic error of the sum rules of the $\xi$-moments, and enable us to calculate higher-order moments. Then we calculated the values of $\xi$-moments up to 10 order, and fitted those values with the LCHO model via the least squares method to determine the behavior of DA $\phi_{2;\pi}(x,\mu)$. This method avoids the unreliability problem of calculating the higher-order Gegenbauer moments of DA by using the QCD sum rules. The more $\xi$-moments give the stronger constraints on DA's behavior. The research idea proposed in Ref.~\cite{Zhong:2021epq} improves the prediction ability of the QCD sum rules in the study of meson DAs, and has been used to study kaon leading-twist DA~\cite{Zhong:2022ecl} and axial-vector $a_1(1260)$-meson longitudinal leading-twist DA~\cite{Hu:2021lkl}. Inspired by the work of Ref.~\cite{Zhong:2021epq}, we try to calculate the value of more $\xi$-moments with order higher than 10, and analyze several commonly used and relatively simple models in the literature such as TF model, LCHO model, DSE model, AdS/QCD model and PLP model, etc., by using the least squares method.

The remaining parts of this paper are organized as follows. In Sec. II, we give a brief overview of several pionic leading-twist DA models to be analyzed. In Sec. III, we give the sum rules of $n$th $\xi$-moment obtained in Ref.~\cite{Zhong:2021epq} for subsequent discussion. Numerical analysis is given in Sec. IV. Sec. V is reserved for a summary.

\section{a brief review on the pionic leading-twist DA models}\label{Sec.II}

\textbf{TF model} By solving the renormalization group equation, the pionic leading-twist DA at scale $\mu$ can be expanded into a Gegenbauer polynomial series~\cite{Lepage:1979zb,Efremov:1979qk}. Its truncation form truncated from the $N$th term may be the most common phenomenological model, which reads
\begin{eqnarray}
\phi_{2;{\pi}}^{\rm TF}(x,\mu) = 6x(1-x) \left[ 1 + \sum^N_{n=2} a_n^{2;{\pi}}(\mu) C^{3/2}_n(2x-1) \right],
\label{TF}
\end{eqnarray}
where $C^{3/2}_n(2x-1)$ is the $n$th order Gegenbauer polynomial, the coefficient $a_n^{2;{\pi}}(\mu)$ is the corresponding Gegenbauer moments. Due to isospin symmetry, the odd Gegenbauer moments vanish and only the even Gegenbauer moments are left. The Gegenbauer moments $a_n^{2;{\pi}}(\mu)$ can be calculated directly via nonperturbative LQCD~\cite{Martinelli:1987si, Daniel:1990ah, DelDebbio:2002mq, Gockeler:2005jz, Braun:2006dg, Donnellan:2007xr, Arthur:2010xf, Braun:2015axa, Bali:2017ude, Detmold:2021qln, RQCD:2019osh} or QCD sum rules~\cite{Chernyak:1981zz, Chernyak:1983ej, Huang:1984nc, Xiang:1984hw, Johnston:1986zi, Mikhailov:1988nz, Mikhailov:1991pt, Bakulev:2001pa, Khodjamirian:2004ga, Ball:2006wn, Bakulev:1998pf}. Due to the limitations of LQCD and QCD sum rules itself and the stability of the Gegenbauer moments calculated by these two nonperturbative methods decreases sharply with the increase of order $n$~\cite{Zhong:2021epq}, the TF model's ability to describe the dependence on a parton momentum fraction $x$ is difficult to improve. Therefore, the following two forms for the TF model are usually adopted in literature,
\begin{eqnarray}
\phi_{2;{\pi}}^{\rm TF,I}(x,\mu) = 6x(1-x) \left[ 1 + a_2^{2;{\pi}}(\mu) C^{3/2}_2(2x-1) \right],
\label{TF1}
\end{eqnarray}
and
\begin{eqnarray}
\phi_{2;{\pi}}^{\rm TF,II}(x,\mu) &=& 6x(1-x) \left[ 1 + a_2^{2;{\pi}}(\mu) C^{3/2}_2(2x-1) \right. \nonumber\\
&+& \left. a_4^{2;{\pi}}(\mu) C^{3/2}_4(2x-1) \right].
\label{TF2}
\end{eqnarray}

\textbf{LCHO Model} By using the approximate bound-state solution of a hadron in terms of the quark model as the starting point, BHL suggest that the hadronic wave function (WF) can be obtained by connecting the equal-time WF in the rest frame and the WF in the infinite-momentum frame. This is the so-called BHL description\cite{BHL}. Based on the BHL description, the LCHO model for pionic leading-twist DA has been established and improved~\cite{Huang:1994dy, Huang:2004fn, Wu:2011gf, Wu:2012kw, Huang:2013gra, Huang:2013yya, Zhong:2015nxa, Zhong:2014jla, Zhong:2021epq}, which reads~\cite{Zhong:2021epq}
\begin{eqnarray}
\phi_{2;\pi}^{\rm LCHO}(x,\mu_0) &=& \frac{\sqrt{3} A_{2;\pi} \hat{m}_q \beta_{2;\pi}}{2\pi^{3/2}f_\pi} \sqrt{x(1-x)} \varphi_{2;\pi}(x) \nonumber\\
&\times& \left\{ \textrm{Erf}\left[ \sqrt{\frac{\hat{m}_q^2 + \mu_0^2}{8\beta_{2;\pi}^2 x(1-x)}} \right] \right. \nonumber\\
&-& \left. \textrm{Erf}\left[ \sqrt{\frac{\hat{m}_q^2}{8\beta_{2;\pi}^2 x(1-x)}} \right] \right\},
\label{LCHO}
\end{eqnarray}
with the longitudinal distribution function
\begin{eqnarray}
\varphi_{2;\pi}(x) = \left[ x(1-x) \right]^{\alpha_{2;\pi}} \left[ 1 + B_2^{2;\pi} C_2^{3/2}(2x-1) \right],
\label{varphi}
\end{eqnarray}
where $\hat{m}_q$ is the constituent $u$ or $d$ quark mass, $f_\pi$ is the pion decay constant. There are several mass schemes for $\hat{m}_q$ in literature. In Ref.~\cite{Zhong:2021epq}, however, we find that, by fitting the values of $\xi$-moments with the LCHO model~\eqref{LCHO}, the goodness of fit increases as $\hat{m}_q$ decreases. We then take $\hat{m}_q = 150\ {\rm MeV}$ in this paper. In addition, $A_{2;\pi}$ is the normalization constant, $\beta_{2;\pi}$ is a harmonious parameters, ${\rm Erf}(x) = 2\int^x_0 e^{-t^2} dx/{\sqrt{\pi}}$ is the error function. The LCHO model is determined by four model parameters, such as $A_{2;\pi}$, $\beta_{2;\pi}$, $\alpha_{2;\pi}$ and $B_2^{2;\pi}$. There are usually two constraints, such as, the DA's normalization condition provided from the process $\pi \to \mu\nu$ and the sum rule derived from $\pi^0 \to \gamma\gamma$ decay amplitude~\cite{Zhong:2021epq}. Then only two model parameters are independence, and which can be constrained by the $\xi$-moments~\cite{Zhong:2021epq} or experimental data~\cite{Wu:2012kw, Huang:2013gra, Huang:2013yya, Zhong:2015nxa}.

\textbf{DSE model} In Ref.~\cite{Chang:2013pq}, the authors suggest a new form for pionic leading-twist DA by combining the truncation of the series of the Gegenbauer polynomials of order $\alpha$, $C_n^\alpha(2x-1)$, and $[x(1-x)]^{\alpha_-}$ with $\alpha_- = \alpha - 1/2$. Comparing with the TF model based on $\{C_n^{3/2}\}$-basis, its advantage is that one can accelerate the procedure's convergence by optimising $\alpha$, reduce non-zero coefficients and reduce the introduced spurious oscillations. The parameters of DSE model are determined by using the first $50$ $x$-moments, $\left<x^n\right> = \int^1_0 dx x^n \phi_{2;\pi}(x,\mu)$, obtained by solving the corresponding gap and Bethe-Salpeter Equations~\cite{Chang:2013pq}. However, the results depend on the kernels. There are obvious differences between the behaviors with the rainbow-ladder and dynamical-chiral-symmetry-breaking-improved kernels~\cite{Chang:2013pq}. Based on Ref.~\cite{Chang:2013pq}, the DSE model for pionic leading-twist DA reads
\begin{eqnarray}
\phi^{\rm DSE}_{2;\pi}(x) = \mathcal{N} [x(1-x)]^{\alpha^{\rm DSE}_-} \left[ 1 + a_2^{\rm DSE} C_2^{\alpha^{\rm DSE}}(2x-1) \right],
\label{DS}
\end{eqnarray}
where $\alpha^{\rm DSE}_- = \alpha^{\rm DSE} - 1/2$, $\mathcal{N}$ is the normalization constant. With the normalization condition of pionic leading-twist DA, $\mathcal{N} = 4^{\alpha^{\rm DSE}} \Gamma(\alpha^{\rm DSE} + 1) / \left[ \sqrt{\pi} \Gamma(\alpha^{\rm DSE} + 1/2) \right]$.

\textbf{AdS/QCD model} The light-front (LF) holographic AdS/QCD has been developed more than ten years ago~\cite{deTeramond:2005su, Brodsky:2006uqa, Brodsky:2007hb, Brodsky:2008pf, deTeramond:2008ht}, which is a semiclassical first approximation to strongly coupled QCD. With this approach, there are holographic duality between the hadron dynamies in physical four-dimensional spacetime at fixed LF time $\tau = x^+ = x^0 + x^3$ and the dynamics of a gravitational theory in five-dimensional anti-de Sitter (AdS) space. Based on the LF holographic AdS/QCD, Ref.~\cite{Chang:2016ouf} suggests two forms for pionic leading-twist DA, such as
\begin{eqnarray}
\phi^{\rm AdS,I}_{2;\pi}(x,\mu) &=& \frac{\sqrt{3}}{\pi f_\pi} \int^{|\textbf{k}_\perp| < \mu} \frac{d^2 \textbf{k}_\perp}{(2\pi)^2} \frac{N_1}{(x\bar{x})^{1/2}} \nonumber\\
&\times& \hat{m}_q \psi(z,\textbf{k}_\perp),
\label{ADSI}
\end{eqnarray}
and
\begin{eqnarray}
\phi^{\rm AdS,II}_{2;\pi}(x,\mu) &=& \frac{\sqrt{3}}{\pi f_\pi} \int^{|\textbf{k}_\perp| < \mu} \frac{d^2 \textbf{k}_\perp}{(2\pi)^2} \frac{N_2}{(x\bar{x})^{1/2}} \nonumber\\
&\times& \left( \hat{m}_q + x\bar{x} \widetilde{m}_\pi \right) \psi(z,\textbf{k}_\perp),
\label{ADSII}
\end{eqnarray}
with the radial WF
\begin{eqnarray}
\psi^{\rm AdS}_{2;\pi}(x,\textbf{k}_\perp) = \frac{4\pi}{\sqrt{\lambda}} \frac{1}{\sqrt{x\bar{x}}} e^{-\frac{\textbf{k}^2_\perp}{2\lambda x\bar{x}}} e^{-\frac{\hat{m}_q^2}{2\lambda x\bar{x}} }.
\label{LFWFFT}
\end{eqnarray}
In which, $\sqrt{\lambda}$ is the mass scale parameter, $\textbf{k}_\perp$ is the transverse momentum, $N_1$ and $N_2$ are the normalization constants and can be obtained with the normalization condition for $\phi^{\rm AdS,I}_{2;\pi}(x,\mu)$ and $\phi^{\rm AdS,II}_{2;\pi}(x,\mu)$ respectively, $\widetilde{m}_\pi = \frac{\hat{m}_q^2 + \textbf{k}_\perp^2}{x(1-x)}$ is the invariant mass of $q\bar{q}$ pair in the pseudoscalar meson~\cite{Chang:2018aut}. The difference between $\phi^{\rm AdS,I}_{2;\pi}(x,\mu)$ and $\phi^{\rm AdS,II}_{2;\pi}(x,\mu)$ is due to that the latter is constructed by further considering the impact of the Dirac structure like $\not\! p \gamma_5$ on spin WF. Substituting Eq.~\eqref{LFWFFT} into Eqs.~\eqref{ADSI} and \eqref{ADSII}, after integrating over the transverse momentum $\textbf{k}_\perp$, the explicit forms of $\phi^{\rm AdS,I}_{2;\pi}(x,\mu)$ and $\phi^{\rm AdS,II}_{2;\pi}(x,\mu)$ can be obtained as
\begin{eqnarray}
\phi^{\rm AdS,I}_{2;\pi}(x,\mu) &=& \frac{2\sqrt{3}N_1\sqrt{\lambda}}{\pi f_\pi} \hat{m}_q \exp \left( -\frac{\hat{m}_q^2}{2\lambda x\bar{x}} \right) \nonumber\\
&\times& \left[ 1 - \exp \left( -\frac{\mu^2}{2\lambda x\bar{x}} \right) \right],
\end{eqnarray}
and
\begin{eqnarray}
\phi^{\rm AdS,II}_{2;\pi}(x,\mu) &=& \frac{2\sqrt{3}N_2\sqrt{\lambda}}{\pi f_\pi} \left( x\bar{x} \right)^{1/2} \left\{ \exp \left( - \frac{\hat{m}_q^2 + \mu^2}{2\lambda x\bar{x}} \right) \right. \nonumber\\
&\times& \left[ \hat{m}_q \exp \left( \frac{\mu^2}{2\lambda x\bar{x}} \right) \left( 1 + (x\bar{x})^{-1/2} \right) \right. \nonumber\\
&-& \left. \hat{m}_q (x\bar{x})^{-1/2} - \sqrt{\hat{m}_q^2 + \mu^2} \right] \nonumber\\
&+& \sqrt{\frac{\pi}{2}} \sqrt{\lambda} \left( x\bar{x} \right)^{1/2} \left[ {\rm Erf} \left( \sqrt{\frac{\hat{m}_q^2 + \mu^2}{2\lambda x\bar{x}}} \right) \right. \nonumber\\
&-& \left.\left. {\rm Erf} \left( \sqrt{\frac{\hat{m}_q^2}{2\lambda x\bar{x}}} \right) \right]  \right\}.
\end{eqnarray}
As pointed out in Refs.~\cite{Chang:2016ouf, Brodsky:2014yha}, the light-quark mass $\hat{m}_q$ in AdS/QCD model is the effective quark mass from the reduction of higher Fock states as functionals of the valence states. Then we do not fix it as in LCHO model. Here, $\hat{m}_q$ together with $\sqrt{\lambda}$ should be undetermined parameters of AdS/QCD model, and can be extracted from observable measurements~\cite{Chang:2016ouf, Brodsky:2014yha, Ahmady:2016ufq}.

\textbf{PLP model} By calculating on 35 different Coordinated Lattice Simulations ensembles with $N_f = 2+1$ flavors of dynamical Wilson-clover fermions, Ref.~\cite{RQCD:2019osh} presents the lattice determination for the first Gegenbauer moments of the pionic leading-twist DA, and suggest a model for pion leading-twist DA based on a simple power-law parametrization,
\begin{eqnarray}
\phi_{2;\pi}^{\rm PLP}(x) = \frac{\Gamma[2\alpha + 2]}{\Gamma[\alpha + 1]^2} x^\alpha (1-x)^\alpha.
\label{PLP}
\end{eqnarray}
Obviously, the PLP model $\phi_{2;\pi}^{\rm PLP}(x)$ satisfies the normalization condition due to the coefficient $\Gamma[2\alpha + 2]/\Gamma[\alpha + 1]^2$. Recently, the PLP model is also used in Ref.~\cite{Li:2022qul} to fit the first 18 Gegenbauer moments by solve the dispersion relation as an inverse problem in the inverse matrix method~\cite{Li:2021gsx} under the framework of the QCD sum rules method.

\section{sum rules for the $\xi$-moments of pionic leading-twist DA}\label{Sec.III}

Considering the fact that the sum rule of the zeroth order $\xi$-moment of pionic leading-twist DA cannot be normalized in the whole Borel parameter region, we suggest the following form for the $n$th order $\xi$-moment in Ref.~\cite{Zhong:2021epq},
\begin{eqnarray}
\langle\xi^n\rangle_{2;\pi}  = \frac{\langle\xi^n\rangle_{2;\pi} \langle \xi^0\rangle_{2;\pi}}{\sqrt{\langle \xi^0\rangle_{2;\pi}^2}}.
\label{xin}
\end{eqnarray}
In which, the numerator is from the following sum rules,
\begin{widetext}
\begin{align}
&\frac{\langle\xi^n\rangle_{2;\pi} \langle \xi^0\rangle_{2;\pi} f_{\pi}^2}{M^2 e^{m_\pi^2/M^2}} = \frac{3}{4\pi^2} \frac{1}{(n+1)(n+3)}  \Big( 1 - e^{-s_\pi/M^2} \Big) + \frac{(m_d + m_u) \langle\bar{q}q \rangle }{(M^2)^2}
~+~ \frac{\langle \alpha_sG^2\rangle }{(M^2)^2}~ \frac{1 + n\theta(n-2)}{12\pi(n+1)}
\nonumber\\[1ex]
&\qquad\qquad~- \frac{(m_d + m_u)\langle  g_s\bar{q}\sigma TGq \rangle }{(M^2)^3}\frac{8n+1}{18} + \frac{\langle g_s\bar{q}q\rangle ^2}{(M^2)^3} \frac{4(2n+1)}{81} - \frac{\langle g_s^3fG^3\rangle }{(M^2)^3}\frac{n \theta(n-2)}{48\pi^2} + \frac{\langle g_s^2\bar{q}q\rangle ^2}{(M^2)^3} \frac{2+\kappa^2}{486\pi^2}
\nonumber\\[1ex]
&\qquad\qquad~ \times \Big\{\!-2(51n+ 25)\Big(\!-\ln \frac{M^2}{\mu^2} \Big) + 3(17n+35) + \theta(n-2)\Big[ 2n  \Big(\!-\ln \frac{M^2}{\mu^2} \Big) + \frac{49n^2 +100n+56}n
\nonumber\\
&\qquad\qquad~ - 25(2n+1)\Big[ \psi\Big(\frac{n+1}{2}\Big) - \psi\Big(\frac{n}{2}\Big) + \ln4 \Big]   \Big] \Big\}.
\label{xinxi0}
\end{align}
\end{widetext}
The denominator in Eq.~\eqref{xin} can be obtained by taking $n = 0$ in Eq.~\eqref{xinxi0}, i.e.,
\begin{align}
&\frac{\langle \xi^0\rangle_{2;\pi} ^2 f_{\pi}^2}{M^2 e^{m_\pi^2/M^2}}
= \frac{1}{4\pi^2} \Big( 1 - e^{-s_\pi / M^2} \Big) + (m_d + m_u)~\frac{ \langle  \bar{q}q \rangle }{(M^2)^2} \nonumber\\
& + \frac{\langle \alpha_sG^2\rangle }{(M^2)^2} \frac{1}{12\pi} - \frac{1}{18}(m_d + m_u) \frac{ \langle  g_s\bar{q}\sigma TGq \rangle }{(M^2)^3} + \frac{4}{81} \frac{\langle g_s\bar{q}q\rangle ^2}{(M^2)^3} \nonumber\\
& + \frac{\langle g_s^2\bar{q}q\rangle ^2}{(M^2)^3} \frac{2+\kappa^2}{486\pi^2} \Big[\!-50 \Big(\!-\ln \frac{M^2}{\mu^2} \Big) + 105 \Big].
\label{xi0xi0}
\end{align}
In Eqs.~\eqref{xinxi0} and \eqref{xi0xi0}, $m_\pi$ is the mass of pion, $M$ is the Borel parameter, $m_{u(d)}$ is the current $u(d)$ quark mass, $s_\pi \simeq 1.05{\rm GeV}^2$~\cite{Zhong:2021epq} is the continuum threshold, $\psi(x)$ is the digamma function. In addition, $\langle  \bar{q}q \rangle$, $\langle \alpha_sG^2 \rangle$, $\langle  g_s\bar{q}\sigma TGq \rangle$ and $\langle g_s^3fG^3\rangle$ are the double-quark condensate, double-gluon condensate, quark-gluon mixed condensate and triple-gluon condensate, respectively, $\langle g_s\bar{q}q\rangle ^2$ and $\langle g_s^2\bar{q}q\rangle ^2$ are two four-quark condensates, $\kappa$ is the ratio of double $s$ quark condensate and double $u/d$ quark condensate.

As discussion in Ref.~\cite{Zhong:2021epq}, the sum rules~\eqref{xin} can provide more accurate values for $\xi$-moments due to that which can also eliminate some systematic errors caused by the continuum state, the absence of high-dimensional condensates, and the various input parameters. In particular, we obtain appropriate Borel windows for first five nonzero $\xi$-moments of pionic leading-twist DA by requiring the dimension-six contribution for all $\left<\xi^n\right>_{2;\pi}$ are no more than $5\%$ and the continuum contribution of $\left<\xi^n\right>_{2;\pi}$ are less than $(30,35,40,40,40)\%$ for $n = (2,4,6,8,10)$, respectively~\cite{Zhong:2021epq}. This, in turn, implies that the systematic error caused by the missing higher-dimension condensates and the continuum and exited states under the quark-hadron duality approximation does not increase with the increase of the order $n$. Then, the sum rules~\eqref{xin} indeed alleviates the limitation of system error on the prediction ability of QCD sum rule method for higher-order $\xi$-moments. Inspired by this, we will calculate the $\xi$-moments $\left<\xi^n\right>_{2;\pi} (n = 12, 14, 16, 18, 20)$ in next section. From the criteria to obtain the allowable Borel windows for those $\xi$-moments, one can find that the values of the $\xi$-moments $\left<\xi^n\right>_{2;\pi} (n = 12, 14, 16, 18, 20)$ obtained in next section are credible.

\section{numerical analysis}\label{Sec.IV}

The first five nonzero $\xi$-moments of pionic leading-twist DA have been calculated with sum rules~\eqref{xin} in Ref.~\cite{Zhong:2021epq}, which are
\begin{eqnarray}
\left<\xi^2\right>_{2;\pi} &=& 0.271 \pm 0.013, \nonumber\\
\left<\xi^4\right>_{2;\pi} &=& 0.138 \pm 0.010, \nonumber\\
\left<\xi^6\right>_{2;\pi} &=& 0.087 \pm 0.006, \nonumber\\
\left<\xi^8\right>_{2;\pi} &=& 0.064 \pm 0.007, \nonumber\\
\left<\xi^{10}\right>_{2;\pi} &=& 0.050 \pm 0.006
\label{xin_value1}
\end{eqnarray}
at scale $\mu = 1{\rm GeV}$.

In order to calculate the values of the $\xi$-moments $\left<\xi^n\right>_{2;\pi} (n = 12, 14, 16, 18, 20)$, we take the inputs as the same as those in Ref.~\cite{Zhong:2021epq}, i.e.,
\begin{eqnarray}
m_\pi &=& 139.57039 \pm 0.00017 {\rm MeV}, \nonumber\\
f_\pi &=& 130.2 \pm 1.2 {\rm MeV}, \nonumber\\
m_u &=& 2.16^{+0.49}_{-0.26} {\rm MeV}, \nonumber\\
m_d &=& 4.67^{+0.48}_{-0.17} {\rm MeV}, \nonumber\\
\left<\bar{q}q\right> &=& \left( -289.14^{+9.34}_{-4.47} \right)^3 {\rm MeV}^3, \nonumber\\
\left<g_S \bar{q}\sigma TGq \right> &=& \left( -1.934^{+0.188}_{-0.103} \right) \times 10^{-2} {\rm GeV}^5, \nonumber\\
\left<g_s\bar{q}q\right>^2 &=& \left( 2.082^{+0.734}_{-0.697} \right) \times 10^{-3} {\rm GeV}^6, \nonumber\\
\left<g_s^2\bar{q}q\right>^2 &=& \left( 7.420^{+2.614}_{-2.483} \right) \times 10^{-3} {\rm GeV}^6, \nonumber\\
\left<\alpha_sG^2\right> &=& 0.038 \pm 0.011 {\rm GeV}^4, \nonumber\\
\left<g_s^3 fG^3 \right> &\simeq& 0.045 {\rm GeV}^6, \nonumber\\
\kappa &=& 0.74\pm 0.03,
\label{input}
\end{eqnarray}
at the scale $\mu = 2{\rm GeV}$. For the scale evolutions of above inputs, one can refer to Ref.~\cite{Zhong:2021epq}.

\begin{table}[t]
\caption{The determined Borel windows and the corresponding pionic leading-twist DA $\xi$-moments $\langle\xi^n\rangle_{2;\pi}$ with $n=(12,14,16,18,20)$. Where all input parameters are set to be their central values.}
\begin{tabular}{ l  c  c }
\hline\hline
$n$ & ~~~~~~~~~~~~~~~~~~~~$M^2$~~~~~~~~~~~~~~~~~~~~ & $\langle\xi^n\rangle_{2;\pi}$ \\
\hline
$12$~ & ~$[5.356,6.082]$~ & ~$[0.0333,0.0308]$~ \\
$14$~ & ~$[6.137,6.888]$~ & ~$[0.0271,0.0252]$~ \\
$16$~ & ~$[6.921,7.721]$~ & ~$[0.0227,0.0211]$~ \\
$18$~ & ~$[7.706,8.578]$~ & ~$[0.0194,0.0180]$~ \\
$20$~ & ~$[8.493,9.456]$~ & ~$[0.0169,0.0156]$~ \\
\hline\hline
\end{tabular}
\label{tbw}
\end{table}
\begin{figure}[t]
\centering
\includegraphics[width=0.42\textwidth]{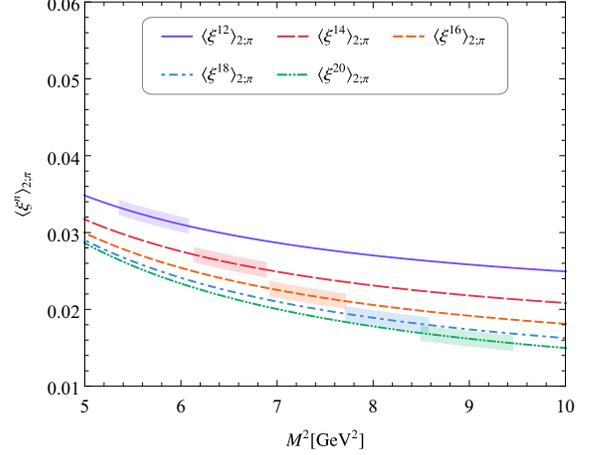}
\caption{The pionic leading-twist DA moments $\langle\xi^n\rangle_{2;\pi}$ with  $n=(12,14,16,18,20)$ versus the Borel parameter $M^2$, where all input parameters are set to be their central values. In which, the shaded band indicate the Borel Windows for $n=(12,14,16,18,20)$, respectively.}
\label{fxinM2}
\end{figure}

Substituting the inputs exhibited in Eq.~\eqref{input} into sum rules~\eqref{xin}, the $\xi$-moments $\left<\xi^n\right>_{2;\pi} (n = 12, 14, 16, 18, 20)$ and the corresponding contributions from continuum states and dimension-six condensates versus Borel parameter $M^2$ can be obtained, respectively. To obtain the allowable Borel windows for $\left<\xi^n\right>_{2;\pi} (n = 12, 14, 16, 18, 20)$, we require the continuum state's contribution and the dimension-six condensate's contribution are less than $40\%$ and $5\%$ for those $\xi$-moments, respectively. The obtained Borel windows and the corresponding values of $\left<\xi^n\right>_{2;\pi} (n = 12, 14, 16, 18, 20)$ are exhibited in Table~\ref{tbw}. Visually, we show the pionic leading-twist DA moments $\langle\xi^n\rangle_{2;\pi}$ with  $n=(12,14,16,18,20)$ versus the Borel parameter $M^2$ in Fig.~\ref{fxinM2}. In the calculation of Table~\ref{tbw} and Fig.~\ref{fxinM2}, all input parameters are set to be their central values.

After taking all uncertainty sources into consideration, the values of the pionic leading-twist DA $\xi$-moments $\langle\xi^n\rangle_{2;\pi}|_{\mu}$ with $n=(12,14,16,18,20)$ can be obtained, such as
\begin{eqnarray}
\left<\xi^{12}\right>_{2;\pi} &=& 0.0408^{+0.0050}_{-0.0049}, \nonumber\\
\left<\xi^{14}\right>_{2;\pi} &=& 0.0346^{+0.0045}_{-0.0045}, \nonumber\\
\left<\xi^{16}\right>_{2;\pi} &=& 0.0301^{+0.0042}_{-0.0041}, \nonumber\\
\left<\xi^{18}\right>_{2;\pi} &=& 0.0266^{+0.0039}_{-0.0038}, \nonumber\\
\left<\xi^{20}\right>_{2;\pi} &=& 0.0239^{+0.0037}_{-0.0036}
\label{xin_value2}
\end{eqnarray}
at scale $\mu = 1{\rm GeV}$.

\begin{table*}[htb]
\caption{Fitting results for different DA models by fitting the values of the first several nonzero pionic leading-twist DA $\xi$-moments exhibited in Eqs.~\eqref{xin_value1} and~\eqref{xin_value2} via the least squares method. }
\begin{tabular}{l  l  l  l  l  l  l }
\hline\hline
~Models~ & ~NS~~~~~~~~~~~~~~~         & ~$3$~        & ~$5$~ & ~$7$~ & ~$9$~ & ~$10$~\\
\hline
~$\phi_{2;\pi}^{\rm TF,I}$~ & ~$a_2^{2;\pi}$~          & ~$0.225$~    & ~$0.247$~ & ~$0.268$~ & ~$0.285$~ & ~$0.293$~ \\
~~ & ~$\chi^2_{\rm min}$~     & ~$0.565723$~ & ~$4.14428$~ & ~$11.4704$~ & ~$21.9983$~ & ~$28.1611$~ \\
~~ & ~$P_{\chi^2_{\rm min}}$~ & ~$0.753624$~ & ~$0.386832$~ & ~$0.0748825$~ & ~$0.00491901$~ & ~$0.000896493$~ \\
\hline
~$\phi_{2;\pi}^{\rm TF,II}$~ & ~$a_2^{2;\pi}$~          & ~$0.205$~    & ~$0.197$~ & ~$0.187$~ & ~$0.177$~ & ~$0.172$~ \\
~~ & ~$a_4^{2;\pi}$~          & ~$0.060$~    & ~$0.098$~ & ~$0.131$~ & ~$0.158$~ & ~$0.170$~ \\
~~ & ~$\chi^2_{\rm min}$~     & ~$0.0162808$~    & ~$0.480741$~ & ~$1.6772$~ & ~$3.78499$~ & ~$5.17049$~ \\
~~ & ~$P_{\chi^2_{\rm min}}$~ & ~$0.898467$~    & ~$0.923102$~ & ~$0.891759$~ & ~$0.804182$~ & ~$0.739208$~ \\
\hline
~$\phi_{2;\pi}^{\rm LCHO}$~ & ~$A_{2;\pi}$~          & ~$5.61391$~    & ~$6.6104$~ & ~$8.78724$~ & ~$8.78198$~ & ~$8.90258$~ \\
~~ & ~$\alpha_{2;\pi}$~          & ~$-0.789$~    & ~$-0.684$~ & ~$-0.504$~ & ~$-0.504$~ & ~$-0.493$~ \\
~~ & ~$B_2^{2;\pi}$~          & ~$-0.163$~    & ~$-0.155$~ & ~$-0.136$~ & ~$-0.135$~ & ~$-0.133$~ \\
~~ & ~$\beta_{2;\pi}$~          & ~$1.56878$~    & ~$1.60137$~ & ~$1.61753$~ & ~$1.62474$~ & ~$1.62855$~ \\
~~ & ~$\chi^2_{\rm min}$~     & ~$0.0522982$~    & ~$0.188917$~ & ~$0.478592$~ & ~$1.0968$~ & ~$1.53822$~ \\
~~ & ~$P_{\chi^2_{\rm min}}$~ & ~$0.819111$~    & ~$0.979358$~ & ~$0.992886$~ & ~$0.993112$~ & ~$0.992055$~ \\
\hline
~$\phi_{2;\pi}^{\rm DSE}$~ & ~$\alpha^{\rm DSE}$~          & ~$1.130$~    & ~$0.948$~ & ~$0.829$~ & ~$0.738$~ & ~$0.703$~ \\
~~ & ~$a_2^{\rm DSE}$~          & ~$0.129$~    & ~$0.048$~ & ~$-0.035$~ & ~$-0.124$~ & ~$-0.166$~ \\
~~ & ~$\chi^2_{\rm min}$~     & ~$0.00559501$~    & ~$0.174471$~ & ~$0.422942$~ & ~$0.694404$~ & ~$0.818127$~ \\
~~ & ~$P_{\chi^2_{\rm min}}$~ & ~$0.940374$~    & ~$0.981061$~ & ~$0.994674$~ & ~$0.998378$~ & ~$0.999157$~ \\
\hline
~$\phi_{2;\pi}^{\rm AdS,I}$~ & ~$N_1$~          & ~$6.24998$~    & ~$5.58365$~ & ~$7.40491$~ & ~$4.99094$~ & ~$8.06215$~ \\
~~ & ~$\sqrt{\lambda}$~          & ~$0.319$~    & ~$0.344$~ & ~$0.305$~ & ~$0.379$~ & ~$0.301$~ \\
~~ & ~$\hat{m}_q$~          & ~$0.076$~    & ~$0.077$~ & ~$0.064$~ & ~$0.075$~ & ~$0.058$~ \\
~~ & ~$\chi^2_{\rm min}$~     & ~$0.0395365$~    & ~$1.13085$~ & ~$3.82526$~ & ~$7.94241$~ & ~$10.3942$~ \\
~~ & ~$P_{\chi^2_{\rm min}}$~ & ~$0.84239$~    & ~$0.769632$~ & ~$0.574839$~ & ~$0.3377$~ & ~$0.238443$~ \\
\hline
~$\phi_{2;\pi}^{\rm AdS,II}$~ & ~$N_2$~          & ~$0.443594$~    & ~$0.405197$~ & ~$0.399993$~ & ~$0.395313$~ & ~$0.392851$~ \\
~~ & ~$\sqrt{\lambda}$~          & ~$1.342$~    & ~$1.212$~ & ~$1.248$~ & ~$1.286$~ & ~$1.300$~ \\
~~ & ~$\hat{m}_q$~          & ~$0.042$~    & ~$0.084$~ & ~$0.082$~ & ~$0.080$~ & ~$0.080$~ \\
~~ & ~$\chi^2_{\rm min}$~     & ~$0.00512763$~    & ~$0.256752$~ & ~$1.20083$~ & ~$2.9544$~ & ~$4.08025$~ \\
~~ & ~$P_{\chi^2_{\rm min}}$~ & ~$0.942914$~    & ~$0.967946$~ & ~$0.944797$~ & ~$0.889188$~ & ~$0.849812$~ \\
\hline
~$\phi_{2;\pi}^{\rm PLP}$~ & ~$\alpha$~          & ~$0.380$~    & ~$0.379$~ & ~$0.370$~ & ~$0.359$~ & ~$0.354$~ \\
~~ & ~$\chi^2_{\rm min}$~     & ~$0.199311$~    & ~$0.219225$~ & ~$0.452207$~ & ~$1.04863$~ & ~$1.46392$~ \\
~~ & ~$P_{\chi^2_{\rm min}}$~ & ~$0.905149$~    & ~$0.994414$~ & ~$0.998372$~ & ~$0.997922$~ & ~$0.997406$~ \\
\hline\hline
\end{tabular}
\label{tFittingResult}
\end{table*}

Now we can fitting the values of $\xi$-moments listed in Eqs.~\eqref{xin_value1} and \eqref{xin_value2} with the models introduced in Sec.~\ref{Sec.II} by adopting the least squares method. For specific fitting procedure, one can refer to Refs.~\cite{Zhong:2021epq, Zhong:2022ecl}. With the brief review on the pionic leading-twist DA models such as TF model, LCHO model, DSE model, AdS/QCD model and PLP model in Sec.~\ref{Sec.II}, we first specify the fitting parameters $\mathbf{\theta}$. They are: $\mathbf{\theta} = (a_2^{2;\pi})$, $(a_2^{2;\pi}, a_4^{2;\pi})$, $(\alpha_{2;\pi}, B_2^{2;\pi})$, $(\alpha^{\rm DSE}, a_2^{\rm DSE})$, $(\sqrt{\lambda}, \hat{m}_q)$, $(\sqrt{\lambda}, \hat{m}_q)$ and $(\alpha)$ for $\phi_{2;\pi}^{\rm TF,I}$, $\phi_{2;\pi}^{\rm TF,II}$, $\phi_{2;\pi}^{\rm LCHO}$, $\phi_{2;\pi}^{\rm DSE}$, $\phi_{2;\pi}^{\rm AdS,I}$, $\phi_{2;\pi}^{\rm AdS,II}$ and $\phi_{2;\pi}^{\rm PLP}$, respectively.

In order to analyze the constraint strength of moments on the model, we will use those pionic leading-twist DA models introduced in Sec.~\ref{Sec.II} to fit the values of the first nonzero three $\xi$-moments, five $\xi$-moments, seven $\xi$-moments, nine $\xi$-moments and ten $\xi$-moments, respectively. That is, we set those values of the $\xi$-moments exhibited in Eqs.~\eqref{xin_value1} and~\eqref{xin_value2} to the form of the following five groups as the fitting samples,
\begin{widetext}
\begin{eqnarray}
{\rm NS}: &3,& \quad \left\{ \left<\xi^2\right>_{2;\pi}, \left<\xi^4\right>_{2;\pi}, \left<\xi^6\right>_{2;\pi} \right\}; \nonumber\\
{\rm NS}: &5,& \quad \left\{ \left<\xi^2\right>_{2;\pi}, \left<\xi^4\right>_{2;\pi}, \left<\xi^6\right>_{2;\pi}, \left<\xi^8\right>_{2;\pi}, \left<\xi^{10}\right>_{2;\pi} \right\}; \nonumber\\
{\rm NS}: &7,& \quad \left\{ \left<\xi^2\right>_{2;\pi}, \left<\xi^4\right>_{2;\pi}, \left<\xi^6\right>_{2;\pi}, \left<\xi^8\right>_{2;\pi}, \left<\xi^{10}\right>_{2;\pi}, \left<\xi^{12}\right>_{2;\pi}, \left<\xi^{14}\right>_{2;\pi} \right\}; \nonumber\\
{\rm NS}: &9,& \quad \left\{ \left<\xi^2\right>_{2;\pi}, \left<\xi^4\right>_{2;\pi}, \left<\xi^6\right>_{2;\pi}, \left<\xi^8\right>_{2;\pi}, \left<\xi^{10}\right>_{2;\pi}, \left<\xi^{12}\right>_{2;\pi}, \left<\xi^{14}\right>_{2;\pi}, \left<\xi^{16}\right>_{2;\pi}, \left<\xi^{18}\right>_{2;\pi} \right\}; \nonumber\\
{\rm NS}: &10,& \quad \left\{ \left<\xi^2\right>_{2;\pi}, \left<\xi^4\right>_{2;\pi}, \left<\xi^6\right>_{2;\pi}, \left<\xi^8\right>_{2;\pi}, \left<\xi^{10}\right>_{2;\pi}, \left<\xi^{12}\right>_{2;\pi}, \left<\xi^{14}\right>_{2;\pi}, \left<\xi^{16}\right>_{2;\pi}, \left<\xi^{18}\right>_{2;\pi}, \left<\xi^{20}\right>_{2;\pi} \right\},
\end{eqnarray}
\end{widetext}
where $\rm NS$ is the abbreviation of the number of samples.

\begin{figure*}[t]
\centering
\includegraphics[width=0.22\textwidth]{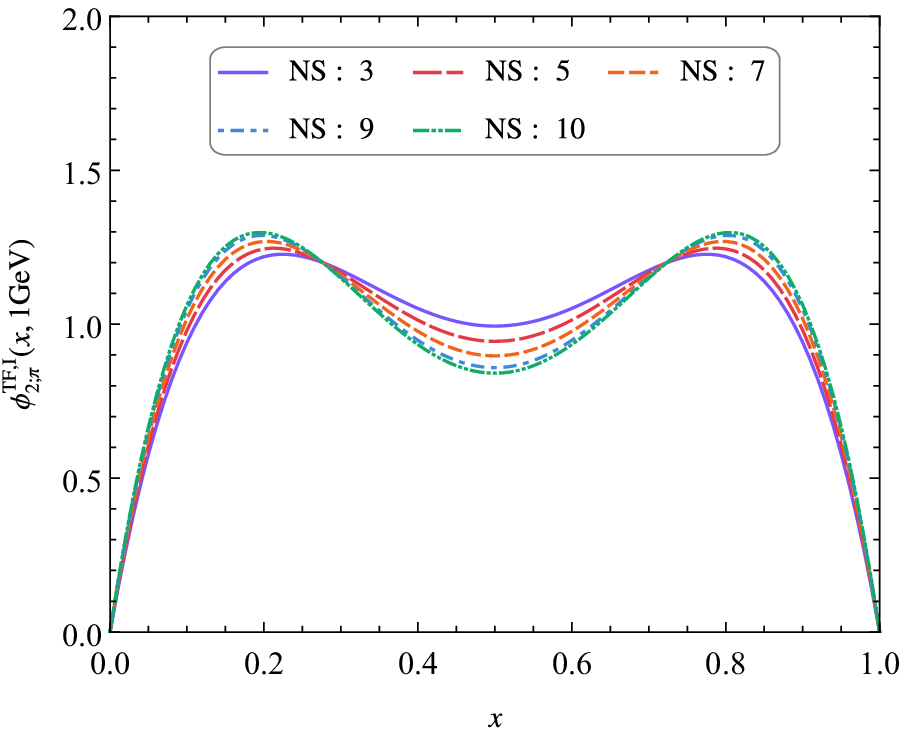}
\includegraphics[width=0.22\textwidth]{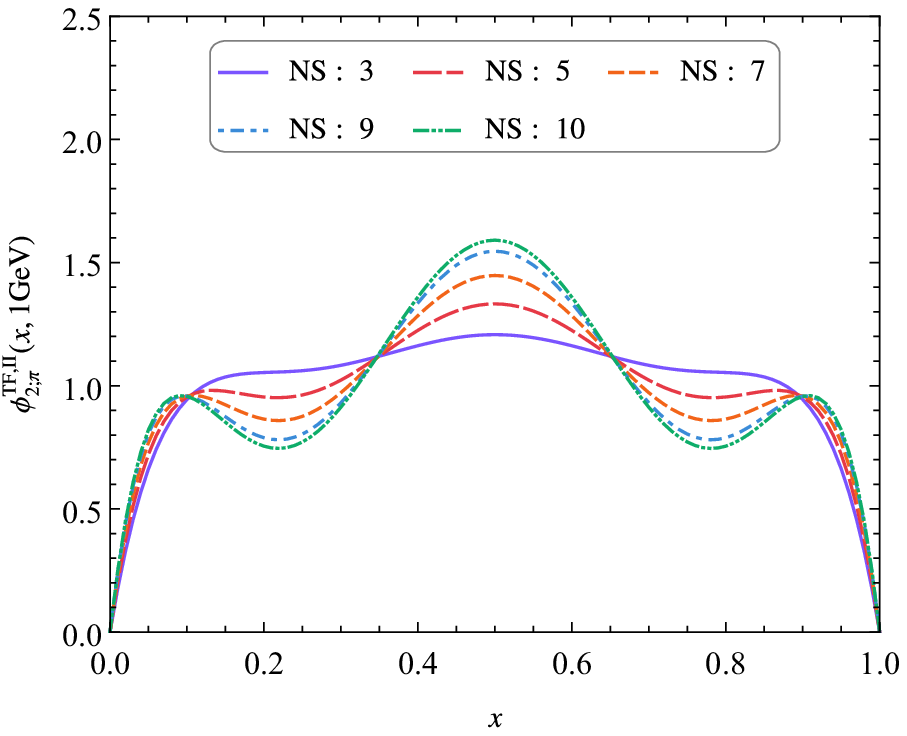}
\includegraphics[width=0.22\textwidth]{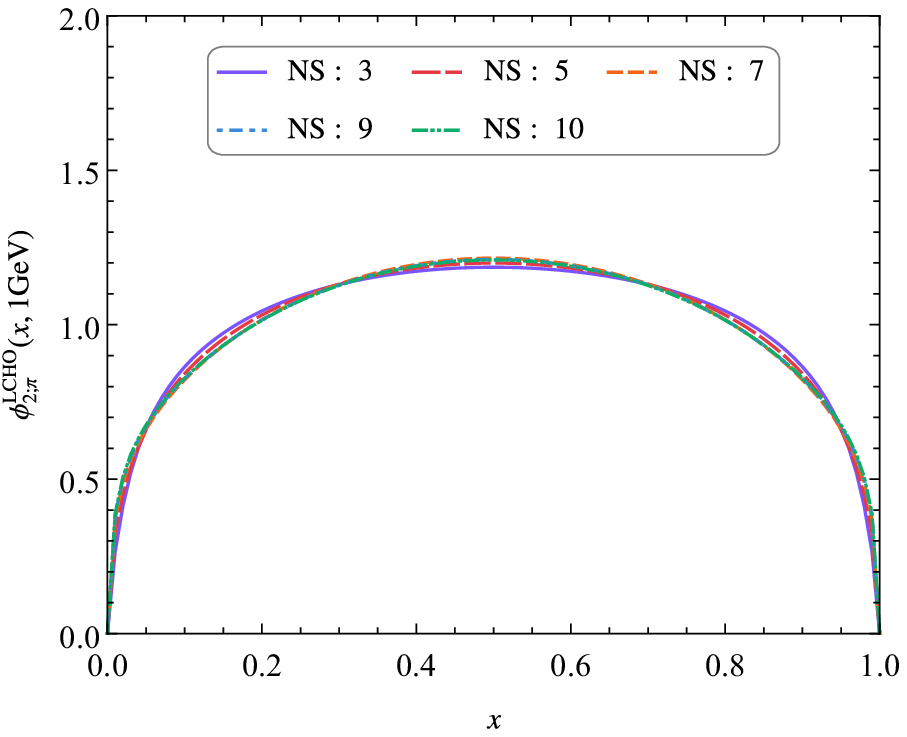}
\includegraphics[width=0.22\textwidth]{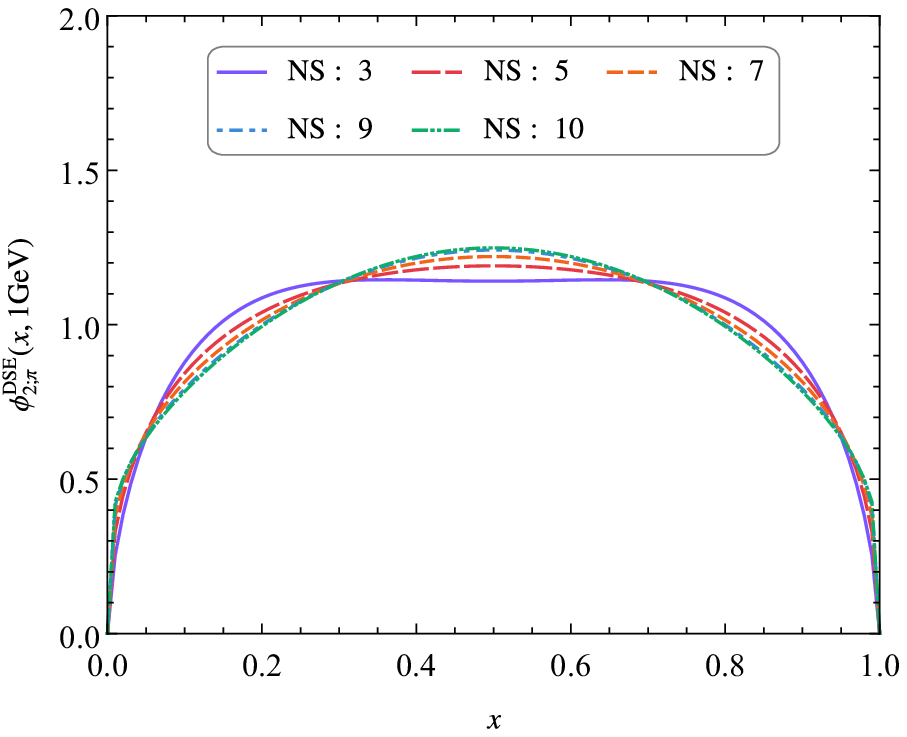}
\includegraphics[width=0.22\textwidth]{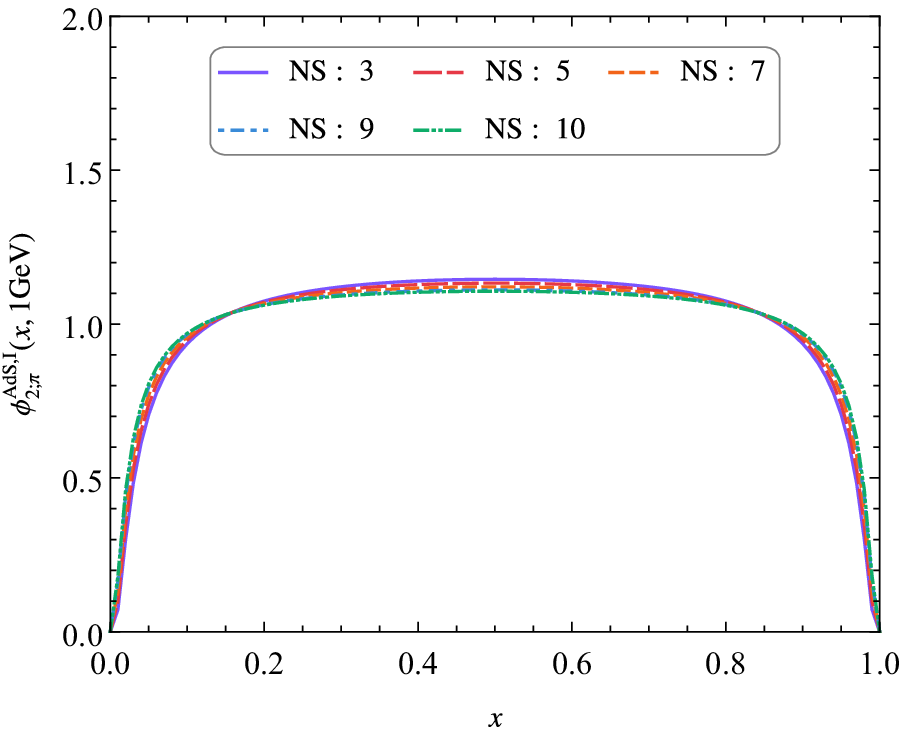}
\includegraphics[width=0.22\textwidth]{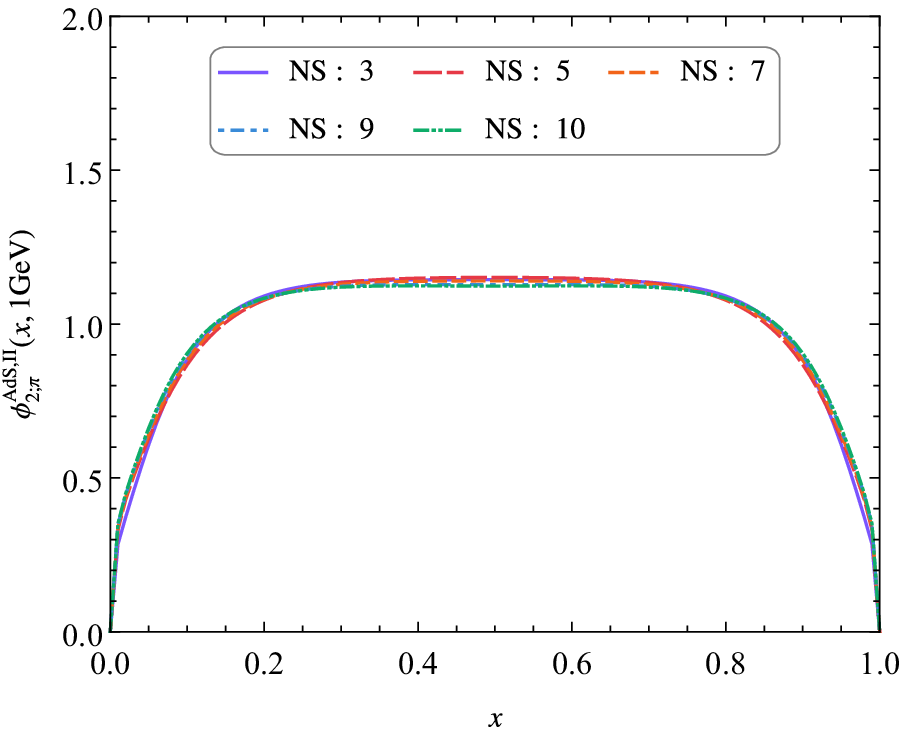}
\includegraphics[width=0.22\textwidth]{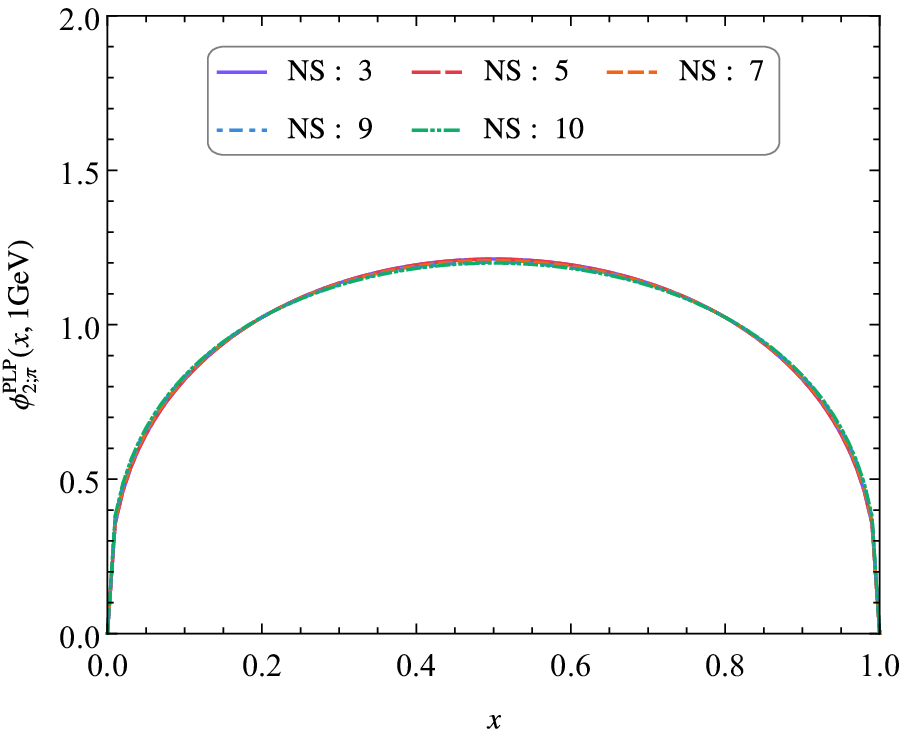}
\caption{The curves of the pionic leading-twist DA models such as TF model, LCHO model, DSE model, AdS/QCD model and PLP model corresponding to the fitting results exhibited in Table~\ref{tFittingResult}.}
\label{fCurves}
\end{figure*}

The fitting results are exhibited in Table~\ref{tFittingResult} and the corresponding model curves are shown in Fig.~\ref{fCurves}. From Table~\ref{tFittingResult} and Fig.~\ref{fCurves}, one can find that,
\begin{itemize}
\item At ${\rm NS} = 10$, $P_{\chi^2_{\rm min}}$ for $\phi_{2;\pi}^{\rm TF,I}$ and $\phi_{2;\pi}^{\rm AdS,I}$ is far less than $1$, $P_{\chi^2_{\rm min}} \simeq 0.74$ for $\phi_{2;\pi}^{\rm TF,II}$, $P_{\chi^2_{\rm min}} \simeq 0.85$ for $\phi_{2;\pi}^{\rm AdS,II}$, while $P_{\chi^2_{\rm min}} > 0.99$ and very closes to $1$ for $\phi_{2;\pi}^{\rm LCHO}$, $\phi_{2;\pi}^{\rm DSE}$ and $\phi_{2;\pi}^{\rm PLP}$. This indicates that the LCHO model, DSE model and PLP model are very consistent with our values of the first 10 nonzero $\xi$-moments exhibited in Eqs.~\eqref{xin_value1} and~\eqref{xin_value2} obtained by the QCD sum rules.

\item The goodness of fit for the models $\phi_{2;\pi}^{\rm TF,I}$ and $\phi_{2;\pi}^{\rm AdS,I}$ decreases with the increases of NS, while the goodness of fit for the models $\phi_{2;\pi}^{\rm LCHO}$, $\phi_{2;\pi}^{\rm DSE}$ and $\phi_{2;\pi}^{\rm PLP}$ is very close to each other with different NS. This indicates that the more moments give the stronger constraints on DA's behavior. Enough moments can easily help us eliminate inappropriate models and select the appropriate model to better describe the correct behavior of DA.

\item The goodness of fit for the model $\phi_{2;\pi}^{\rm AdS,II}$ is better than that for $\phi_{2;\pi}^{\rm AdS,I}$ under ${\rm NS} = 3,5,7,9,10$. This indicates that $\phi_{2;\pi}^{\rm AdS,II}$ by considering the influence of the Dirac structure $\not\! p\gamma_5$ on WF is more consistent with our sum rule results for $\xi$-moments. On the other hand, by fitting first ten nonzero $\xi$-moments with $\phi_{2;\pi}^{\rm AdS,II}$, we obtain the model parameters $\sqrt{\lambda} = 1.300 {\rm GeV}$, $\hat{m}_q = 80 {\rm MeV}$. The value of $\sqrt{\lambda}$ is much larger than the one in literature. The value of $\hat{m}_q$ we obtained is consistent with the value by fitting the pion and kaon decay constants in Ref.~\cite{Chang:2016ouf}, but is larger than the value by fitting the Regge trajectories of pseudoscalar mesons in Ref.~\cite{Brodsky:2014yha}.

\item The curves of $\phi_{2;\pi}^{\rm TF,I}$ and $\phi_{2;\pi}^{\rm TF,II}$ with different NS are significantly different from each other. This shows that the models $\phi_{2;\pi}^{\rm TF,I}$ and $\phi_{2;\pi}^{\rm TF,II}$ are far from enough to describe the behavior of pionic leading-twist DA. Therefore, the DA obtained by calculating the second and/or fourth Gegenbauer moments with LQCD or QCD sum rules and substituting into $\phi_{2;\pi}^{\rm TF,I}$ and $\phi_{2;\pi}^{\rm TF,II}$ in Eqs.~\eqref{TF1} and~\eqref{TF2} are also far from enough to describe the behavior of pionic leading-twist DA. However, the goodness of fit of $\phi_{2;\pi}^{\rm TF,II}$ is better than that of $\phi_{2;\pi}^{\rm TF,I}$ with the same NS. This indicates that by increasing the number of terms in TF, i.e., $N$ in Eq.~\eqref{TF}, the ability of TF model to describe the behavior of pionic leading-twist DA can be increased.

\item The curves of $\phi_{2;\pi}^{\rm LCHO}$ with ${\rm NS = 3,5,7,9,10}$ almost coincide with each other, the same results are for $\phi_{2;\pi}^{\rm AdS,I}$, $\phi_{2;\pi}^{\rm AdS,II}$ and $\phi_{2;\pi}^{\rm PLP}$. This indicates that the LCHO model, AdS/QCD model and PLP model have strong prediction ability for the behavior of pionic leading-twist DA. With first few Gegenbauer moments or $\xi$-moments to constrain the model parameters, the obtained those three models are enough to describe the behavior of pionic leading-twist DA. However, from the goodness of fit for $\phi_{2;\pi}^{\rm AdS,I}$, $\phi_{2;\pi}^{\rm AdS,II}$ exhibited in Table~\ref{tFittingResult}, one can find that those two models are not very consistent with our QCD sum rule results.
\end{itemize}

\section{summary}

In this paper, we calculate the pionic leading-twist DA $\xi$-moments $\left<\xi^n\right>_{2;\pi}(n = 12,14,16,18,20)$ with the sum rule formula of $n$th $\xi$-moment, Eq.~\eqref{xin}, suggested in our previous work~\cite{Zhong:2021epq}. In calculation, the contributions from continuum state and dimension-six condensate are less than $40\%$ and $5\%$ for those five $\xi$-moments. Due to the form of sum rules Eq.~\eqref{xin}, the limitation of system error caused by the missing higher-dimension condensates and the continuum and exited states under the quark-hadron duality approximation on the prediction ability of QCD sum rule method for higher-order $\xi$-moments is alleviated. Then the values of $\xi$-moments $\left<\xi^n\right>_{2;\pi}(n = 12,14,16,18,20)$  are reliable and are exhibited in Eq.~\eqref{xin_value2}. By combining the values of $\left<\xi^n\right>_{2;\pi}(n = 2,4,6,8,10)$ obtained in our previous work~\cite{Zhong:2021epq}, and then taking these values of first ten nonzero $\xi$-moments as the samples, we perform the analysis for several commonly and relative simple models of pionic leading-twist DA, such as, TF model, LCHO model, DSE model, AdS/QCD model and PLP model, with the least squares method. We find that, the TF model is not enough to describe the behavior of pionic leading-twist DA. The LCHO model, AdS/QCD model and PLP model have strong prediction ability for the pionic leading-twist DA. The LCHO model, DSE model and PLP model are very consistent with our values of the first 10 nonzero $\xi$-moments exhibited in Eqs.~\eqref{xin_value1} and~\eqref{xin_value2} obtained by the QCD sum rules.

{\bf Acknowledgments}:
We are grateful to Professor Qin Chang for helpful discussions and valuable suggestions. This work was supported in part by the National Natural Science Foundation of China under Grants No. 11765007, No. 11947406, No. 12147102, No. 11875122, and No. 12175025; the Project of Guizhou Provincial Department of Science and Technology under Grant No. ZK[2021]024; the Project of Guizhou Provincial Department of Education under Grants No. KY[2021]030 and No. KY[2021]003; the Chongqing Graduate Research and Innovation Foundation under Grant No. ydstd1912; the Fundamental Research Funds for the Central Universities under Grant No. 2020CQJQY-Z003; and the Project of Guizhou Minzu University under Grant No. GZMU[2019]YB19.

\end{document}